\documentclass[aip,amsmath,amssymb,reprint]{revtex4-1}
\usepackage{graphicx}
\usepackage{dcolumn}
\usepackage{bm}
\usepackage[utf8]{inputenc}
\usepackage[T1]{fontenc}
\usepackage{mathptmx}
\usepackage{etoolbox}

\makeatletter
\def\@email#1#2{%
 \endgroup
 \patchcmd{\titleblock@produce}
  {\frontmatter@RRAPformat}
  {\frontmatter@RRAPformat{\produce@RRAP{*#1\href{mailto:#2}{#2}}}\frontmatter@RRAPformat}
  {}{}
}%
\makeatother

\begin{document}

\title{Nanoscopic jets and filaments of superfluid $^4$He at zero temperature:\\ a DFT study}

\author{Francesco Ancilotto}
\affiliation{Dipartimento di Fisica e Astronomia ``Galileo Galilei''
and CNISM, Universit\`a di Padova, via Marzolo 8, 35122 Padova, Italy}
\affiliation{ CNR-IOM Democritos, via Bonomea, 265 - 34136 Trieste, Italy }

\author{Manuel Barranco}
\affiliation{Departament FQA, Facultat de F\'{\i}sica,
Universitat de Barcelona, Av.\ Diagonal 645,
08028 Barcelona, Spain.}
\affiliation{Institute of Nanoscience and Nanotechnology (IN2UB),
Universitat de Barcelona, Barcelona, Spain.}

\author{Mart\'{\i} Pi}
\affiliation{Departament FQA, Facultat de F\'{\i}sica,
Universitat de Barcelona, Av.\ Diagonal 645,
08028 Barcelona, Spain.}
\affiliation{Institute of Nanoscience and Nanotechnology (IN2UB),
Universitat de Barcelona, Barcelona, Spain.}

\begin{abstract}
The instability of a cryogenic $^4$He jet exiting through  a small nozzle into vacuum 
leads to the formation of $^4$He drops which are considered 
as ideal matrices for spectroscopic studies of embedded atoms and molecules. 
Here, we present a He-DFT description of  droplet formation resulting 
from jet breaking and contraction of superfluid $^4$He filaments.   
Whereas the fragmentation of long jets closely follows the predictions
of linear theory for inviscid fluids, leading to droplet trains 
interspersed with smaller satellite droplets, the contraction of filaments with an aspect 
ratio larger than a threshold value leads to the nucleation of vortex rings 
which hinder their breakup into droplets. 

\end{abstract}

\date{\today}

\maketitle

\section{Introduction}

Liquid $^4$He droplets at low temperature offer a unique environment 
for molecular spectroscopy\cite{Leh98,Cho06,Cal11} and the study of 
superfluidity on the atomic scale,\cite{Sin89,Kri90,Gre98} including the study of quantum 
vortices.\cite{Gom14,Lan18,Ges19,Oco20}
Usually, $^4$He droplets are produced by expansion of cooled $^4$He gas or 
by instability of a cryogenic $^4$He jet exiting a 
source chamber into vacuum throughout a nozzle, whose temperature and pressure  
determine the appearance of the liquid jet 
and the droplet size and velocity distributions.\cite{Toe04,Sle22} 
$^4$He drops undergo evaporative cooling and
become superfluid, eventually reaching a 
temperature of 0.4 K\cite{Toe04} on a $\mu$s time scale.\cite{Tan18}

Understanding the dynamical properties of liquid $^4$He jets and 
the instabilities leading to their fragmentation  
is a relevant issue in the production and characterization of droplets made of
$^4$He. This unique fluid allows for a large variation of  non-dimensional parameters 
related to the fluid viscosity and the velocity at which it exits the nozzle, 
which characterize its dynamical properties.\cite{Spe20} 
This understanding has also a primary application, namely  
to make available $^4$He drops with the  size and velocity required by the experiments, together
with size and velocity distributions as narrow as possible.
This has led to recent experimental studies on the disintegration of liquid $^4$He jets.\cite{Tan20,Kol22}
Besides, a liquid thread with finite length 
(``filament'' in the following) with no external constraint is expected to contract
trying to minimize its surface energy and eventually reach a spherical liquid drop. However,
the outcome of the process is not always that simple, as an ample body of experiments
and theoretical work on classical fluids has shown in the years. 
We notice that liquid $^4$He filaments are regularly found in the 
experiments.\cite{Spe20,Tan20,Kol22} 

Liquid jets and filaments and their dynamical instabilities are  
well established subjects of study in classical fluids dynamics
because of practical questions and applications on the one hand, 
and because jet dynamics probes many physical properties 
and theoretical approaches on the other hand, see e.g. 
Refs. \onlinecite{Egg08,Cas12,Ant19} and references therein.
Most studies concentrate on viscous fluids because of practical implications. 
The underlying theoretical and numerical challenge is to solve the
Navier-Stokes (NS) equation subject to appropriate boundary conditions.

The effect of  viscosity and surface tension is embodied in the Ohnesorge number $Oh$ defined as
$Oh= \mu/\sqrt{m\rho_0 \gamma R_0}$,
where $m$ is the atom mass, $\rho_0$ the atom density of the fluid, $\gamma$ the surface tension,
$\mu$ the viscosity coefficient, and $R_0$ the radius of the jet or filament.  
Inviscid filaments have been addressed in passing 
by Schulkes,\cite{Sch96} but owing to computational challenges, he could not
simulate extreme interfacial deformations arising in crucial moments of the dynamics, as
during filament breaking and end-pinching, i.e. the formation of two isolated drops from the 
opposite tips of the filaments. 
While it is naturally assumed that solving the NS equation for small enough viscosities the results should be 
nearly indistinguishable from the inviscid limit, see e.g., Refs. \onlinecite{Ant19,Hoe13}, 
a description of superfluid (i.e. inviscid {\it and} irrotational)
jets and filaments is not available in the literature. 
Such study may be of relevance in view of
the aforementioned studies on superfluid $^4$He droplets and it is the 
motivation of the present work. Our goal is to describe,
at the microscopic level, the dynamics of 
contraction and breaking 
of zero temperature superfluid $^4$He nanojets and 
nanofilaments in vacuum using the well-established
$^4$He density functional (He-DFT) approach.\cite{Dal95,Bar06,Anc17,dft-guide}
The He-DFT approach is similar, in the superfluid $^4$He phase, to the Gross-Pitaevskii approach  which
has successfully been applied to the description of cold gases in the superfluid Bose-Einstein condensate phase, in particular in
the study of quantized vortices.\cite{Pit16,Bar16,Tsu09}

Helium density functional and time-dependent density functional (He-TDDFT) methods
have proven to 
be very powerful tools to address superfluid $^4$He samples. 
Within the He-DFT approach, the finite range of 
the helium-helium van der Waals (vdW) interaction is explicitly  incorporated in the
simulations. As a consequence, the liquid-vacuum interface
has a non-zero surface width, which is important in the 
description of nanoscopic $^4$He systems like the jets and filaments studied in the present work.
It also takes into account the finite compressibility of the fluid and therefore the possibility of having
density excitations (ripplons, phonons and rotons) is naturally 
incorporated into the simulations. It also considers the possibility of atom evaporation 
from the $^4$He sample during the real-time dynamics,\cite{Gar22}
which however has been found to be a negligible effect in the present study.
 
The He-DFT approach adds to the classical viscous fluids or molecular dynamics 
descriptions the possibility of disclosing purely superfluid effects 
in the dynamics, in particular quantized vortex nucleation. 
It has been recognized that a retracting viscous liquid filament may escape from end-pinching
through the creation of vortex rings for Ohnesorge numbers in the $0.002 < Oh < 0.1$ range.\cite{Hoe13}
Here we show that the same happens in the zero viscosity, irrotational superfluid case. 
  
Due to the computational burden associated with fully three-dimensional
He-DFT simulations as the ones discussed here, 
we address jets and filaments of nanoscopic size.
Studies on the breakup of liquid nanojets are available in the literature; atomistic molecular dynamics simulations 
on the formation, stability and breakup of viscous fluids have been carried out.\cite{Mos00} 
To our knowledge,
no simulations of breakup of superfluid nanojets and filaments have been published so far.

This work is organized as follows. In Sect. II we briefly present the He-DFT approach used in this work.
In Sect III.A we discuss the results for the dynamics of $^4$He jets, focusing on the 
conditions leading to fragmentation,
and in Sect. III.B we study the contraction and possible break-up of $^4$He filaments
with finite length.
A summary with some concluding remarks is presented in Sect. IV. 
In addition to the main text, we provide in the supplementary material  
the real-time dynamics of the $^4$He jets and filaments addressed in this
paper. This multimedia material  constitutes an important part of this work, since  
it helps capture physical details which would otherwise escape the written account.

\section{Theoretical Approach}
 
Density functional theory for liquid helium is a phenomenological approach
which constitutes a good compromise between accuracy and feasibility. 
The parameters of the functional have been
adjusted to reproduce various properties of the bulk superfluid 
such as equilibrium density, energy per atom and compressibility, as well
as the main features of the dispersion relation of the elementary excitations of superfluid $^4$He.\cite{Dal95}
A detailed description of the method can be found in Refs. \onlinecite{Bar06,Anc17,dft-guide}.

Within He-DFT, the energy of
a $N$-atom sample is written as a functional of the $^4$He atom
density $\rho({\mathbf r})$ as
\begin{equation}
E[\rho] = T[\rho] + E_c[\rho] =
\frac{\hbar^2}{2m} \int d {\mathbf r} |\nabla \Psi({\mathbf r})|^2 
+  \int d{\mathbf r} \,{\cal E}_c[\rho]
\label{eq1}
\end{equation}
where the first term is the kinetic energy, 
$m$ is the mass of the $^4$He atom and
$\Psi({\mathbf r})$ is the effective wave function (or order
parameter) of the superfluid such that 
$\rho({\mathbf r})=|\Psi({\mathbf r})|^2$ with $\int d{\bf r}|\Psi({\bf r})|^2 =
N$. The functional ${\cal E}_c(\rho)$ we have used contains the
He-He interaction term within the Hartree approximation and additional terms
describing non-local correlation effects.\cite{Anc05}

The equilibrium configuration of the system is obtained by solving, 
using an imaginary-time relaxation method,\cite{dft-guide} the
Euler-Lagrange equation
\begin{equation}
\left\{-\frac{\hbar^2}{2m} \nabla^2 + \frac{\delta {\cal
    E}_c}{\delta \rho}  \right\}\Psi \equiv {\cal H}[\rho] \,\Psi  = \zeta \,\Psi
 \label{eq2}
\end{equation}
where  $\zeta$ is the $^4$He chemical potential corresponding to the
number of He atoms in the sample.

Minimizing the action associated to Eq. (\ref{eq1}) leads to
the He-TDDFT equation
\begin{equation}
i \hbar \frac {\partial \Psi}{\partial t}= \left\{-\frac{\hbar^2}{2m} \nabla^2 + 
\frac{\delta {\cal E}_c}{\delta \rho}  \right\}\Psi 
 \equiv {\cal H}[\rho] \,\Psi   
\label{eq3}
\end{equation}
from which one can simulate the real-time evolution of the system.

The above equations have been solved using the $^4$He-DFT-BCN-TLS 
computing package,\cite{Pi17} see  Refs.~\onlinecite{Anc17} and \onlinecite{dft-guide} 
and references therein for additional details.
Briefly, we work in cartesian coordinates, with the effective wave 
function $\Psi(\mathbf{r},t)$ defined at the nodes of a 3D grid inside a
calculation box.
Periodic boundary conditions (PBC) are imposed which allow to use the Fast Fourier 
Transform\cite{Fri05} to efficiently compute the convolutions needed to obtain the DFT
mean field ${\cal H}[\rho]$. The differential operators 
in ${\cal H}[\rho]$ are approximated by 13-point
formulas.
Eqs. (\ref{eq2}-\ref{eq3}) have been solved  using a
space-step of 1.2 \AA{}, and
the time-dependent Eq. (\ref{eq3}) has been
numerically integrated using a Hamming predictor-modifier-corrector
initiated by a fourth-order Runge-Kutta-Gill algorithm\cite{Ral60}
with a time-step of 2~fs. This time-step has been found to keep the energy of the jet and filaments 
properly conserved during the dynamics, as it corresponds to non-dissipative processes. We have also checked
that the   jet configurations obtained in the course of the dynamics are robust against reasonable changes of the chosen 
space-step.

\section{Results}

We have considered jets and filaments of sharp radius $R_0= 21.5$~\AA, 
defined as the radius at which the density equals $\rho_0/2$,
$\rho_0$ being the liquid $^4$He atom density at zero temperature and pressure.  

\subsection{$^4$He nanoscopic jet}

The physics of classical liquid jets has been reviewed by Eggers and Villermaux.\cite{Egg08}
The thinning and breakup of a liquid jet is mainly determined by 
surface tension effects.
The stability of an infinite fluid cylinder of radius $R_0$ was studied by Plateau,\cite{Pla57}
showing that it exists in an unstable equilibrium, and
any perturbation with wavelength $\lambda $ greater than $2\pi R_0$
is unstable and allows the surface tension to break up the cylinder into droplets, thus 
decreasing the surface energy of the system.
Lord Rayleigh later showed\cite{Ray79} that for an inviscid liquid the 
fastest growing mode occurs when the wavelength
of the axial undulation that ultimately leads to the fragmentation
of the jet into droplets is equal to  
$\lambda_c = 9.01\, R_0$ (Rayleigh-Plateau instability).
When the jet breaks up, one or more small satellite drops -resulting from 
the necks breaking- may form between
the larger droplets.

The characteristic times for jet instability and breakup are set
by the capillary time $\tau_c$ defined as
\begin{equation}
\tau_c = \sqrt{\frac{m \rho_0 R_0^3}{\gamma}}
\end{equation}
with $\gamma$ being the   surface tension of the liquid.
In the case of $^4$He we have $m=4.325 \times  10^{13}$ K,   
$\rho_0= 0.021836$ \AA$^{-3}$  and $\gamma= 0.274$
K \AA$^{-2}$. Hence, $\tau_c(R_0=21.5 {\rm \AA})=61.7 \,{\rm ps}$.

It is customary to define the aspect ratio as 
$ \Gamma =\tilde{L}/R_0$ where $\tilde{L}=L/2$ is the
half-length of the jet. Here $L$ coincides with the length of the simulation cell in
the jet direction. From the linearized fluid dynamics equations for 
an inviscid and incompressible fluid, a critical value $\Gamma_c$
is predicted to trigger jet fragmentation.\cite{Ray79} It corresponds to 
the mode with wavelength $\lambda_c =2\pi/k$, where $k$ is such that $\omega (k)$
is maximum. Here\cite{Egg08}
\begin{equation}
\omega ^2(k)= \left[ \xi\,\frac{I_1(\xi)}{I_0(\xi)}(1-\xi^2)\right]\,\frac{1}{\tau_c^2}
\label{omk}
\end{equation}
where $I_0(x)$ and $I_1(x) = d I_0(x)/dx$ 
are  modified Bessel functions of the first kind and 
$\xi \equiv kR_0$. From the maximum of $\omega $ one finds $kR_0=0.697$ and thus
\begin{equation}
\Gamma_c=\pi/0.697=4.505
\label{gammac}
\end{equation}
In correspondence with it one has
\begin{equation}
\omega_{max}=0.343\,\sqrt{\gamma /(m \rho_0 R_0^3)}= \frac {0.343}{\tau _c}
\label{omm}
\end{equation}
\medskip

\begin{figure}
    \centering
    \includegraphics[width=1.0\linewidth,clip]{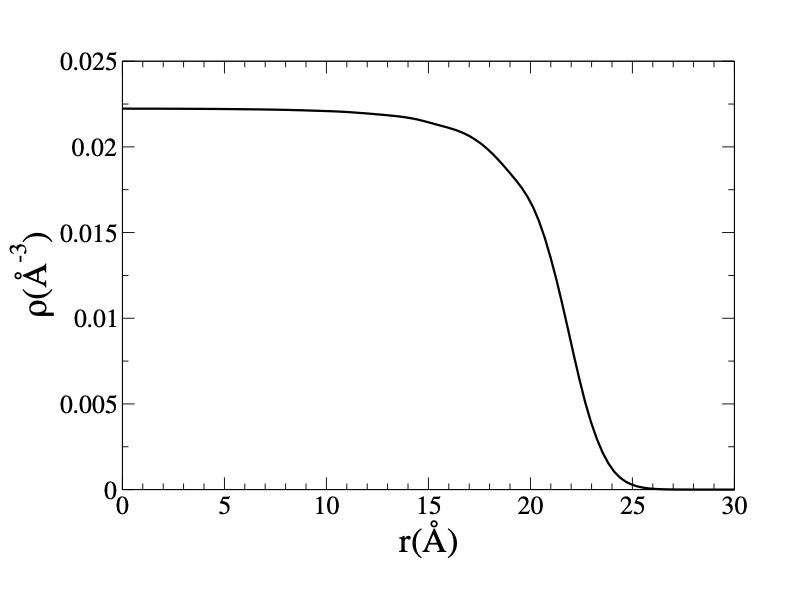}
    \caption{Density profile in the radial direction of a cylinder of radius $R_0=21.5$ \AA{} 
    representing a $^4$He nanojet. }
    \label{fig1}
\end{figure}

Similarly to what occurs in classical liquid jets,
we have shown that the He-TDDFT approach yields the Rayleigh-Plateau instability
for the superfluid nanojet when it is subject to a 
perturbation with the right wavelength. Since the evolution takes place in vacuum, i.e. in the 
absence of ambient gas embedding the jet, the velocity of the jet itself
is not playing any role and therefore we perform our simulations
in a reference frame where the jet is at rest (comoving frame). 
To this end,
we simulate the jet by a cylindrical filament in a simulation
box subject to PBC along the cylinder axis (the $x$-axis in the following).
Its equilibrium structure has been obtained by solving Eq. (\ref{eq2}); 
a plot of the jet density profile in the transverse  direction is shown
in Fig. \ref{fig1}. The jet displays a bulk region of fairly constant 
density (slightly higher than the bulk $^4$He density $\rho_0$ due to the compressive effect
exerted by the surface tension on the lateral surface of the cylinder), 
delimited by a surface with a finite width. As mentioned above, the radius of the cylinder 
$R_0$ is defined as the distance from the symmetry axis of the point 
where $\rho(r)= \rho_0/2$.

We have verified by imaginary-time dynamics that 
the cylinder is indeed unstable against a small initial 
axial perturbation with the proper wavelength.
To do so, we consider a (periodically repeated) 
cylinder made of $N=12076$ $^4$He atoms and length $L=387.2$ \AA.
We have first found the equilibrium geometry with a resulting radius
$R_0 =  21.5$\,\AA{}. Therefore 
the aspect ratio of the cylinder is $\Gamma=\tilde{L}/R_0=9.01$,  i.e., 
twice the critical aspect ratio Eq. (\ref{gammac}); in this way 
the axial undulation caused by the mode with the Rayleigh-Plateau wavelength
$\lambda_c$
will produce two necks along the jet inducing the fragmentation into 
two droplets, as shown in the following.

Next, we performed an imaginary-time dynamics
starting from a configuration corresponding to a a slightly perturbed axially symmetric cylinder
of radius $R_0$, where the initial density profile is given by:
\begin{equation}
\rho(\mathbf{r})=\frac{\rho _0}{exp\{[\sqrt{y^2+z^2}-R(x)]/0.5\}+1}
\end{equation}
with
\begin{equation}
R(x)=R_0[1-\epsilon \cos(4\pi x/L)]
\label{eqrad}
\end{equation}
and $\epsilon \ll 1$.
With our choice of the length $L$ and radius $R_0$, the wavelength of the
resulting density modulation is precisely equal to $\lambda _c=9.01\,R_0$.
The form in Eq. (\ref{eqrad}) ensures that the perturbed density is normalized so to 
have the same number of atoms as the unperturbed cylinder.
If $\delta_0 = \epsilon \,R_0$, the maximum excursion of the radius along the cylinder axis is thus $R=R_0\pm \delta_0$.

The total energy of the axially perturbed cylinder turns out to be lower than that 
of the unperturbed cylinder, i.e. the system
is energetically unstable toward a deformation leading to fragmentation.
Starting from this configuration, we have performed an 
imaginary-time relaxation during which two necks develop eventually leading,
as they shrink to zero, to two identical spherical droplets as the lowest energy state.

Next, we have studied, by solving the He-TDDFT  Eq. (\ref{eq3}),
the actual real-time dynamics of the fragmentation process, 
starting from the axially perturbed cylindrical jet.
Following Ref. \onlinecite{Dri13},
the perturbation is applied both to the density, as described above, and to the 
axial velocity of the jet as well, using the linearized solution 
of the Rayleigh-Plateau instability
\begin{equation}
v=v_0\sin(4\pi x/L)
\end{equation}
where $v_0=2\,\delta_0\,v_{max}/R_0$.
Notice that the radial perturbation is symmetric in the origin, 
whereas the velocity fluctuation is anti-symmetric.
Here 
$v_{max}=\omega _{max}R_0/\xi_{max}$ 
is calculated from Eq.(\ref{omk}) using $\xi=\xi_{max}=0.697$, giving $v_{max}\sim 17$\,m/s.
Our starting value for the perturbation amplitude is $\delta _0=0.452$ \AA, 
corresponding to the choice $\epsilon = 0.021$ in Eq. (\ref{eqrad}).

In order to apply this velocity field to the superfluid jet, we multiply the initial axially perturbed cylinder wave function 
$\Psi(\mathbf{r}) = \rho^{1/2}(\mathbf{r})$ by the phase $e^{i\phi}$ with
\begin{equation}
\phi =-2\frac {\delta _0}{R_0}v_{max}\left(\frac {L/2}{2\pi}\right)\cos(4\pi x/L)
\end{equation}
and proceed with the real-time evolution.

\begin{figure*}
    \centering
    \includegraphics[width=0.9\linewidth,clip]{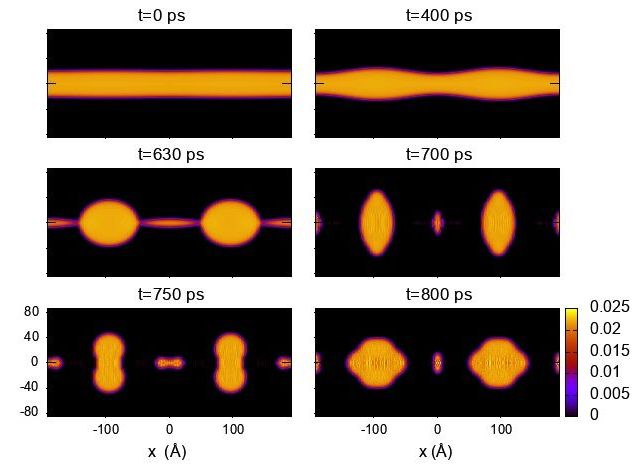}
    \caption{Breaking dynamics of a cylinder subject 
to an axial perturbation of wavelength $\lambda =2\pi R_0/0.697$.
The color bar shows the atom density in units of \AA$^{-3}$.}
    \label{fig2}
\end{figure*}

Figure \ref{fig2} shows snapshots of the jet density 
on a symmetry plane containing the cylinder
axis during the real-time dynamics. 
It can be seen that,
starting from the perturbed cylinder, undulations whose amplitude increases with time 
appear along the jet. The instability is
caused by the fact that the Laplace pressure increases in
constricted regions, driving out the fluid and hence reducing
further the neck radius.
The jet evolves into density bulges 
connected by thin threads.  
Threads eventually break up and isolated drops appear instead.
Figure \ref{fig2} also shows that the threads between drops contract and develop small end droplets 
(see the panel at $t=700$ ps) that subsequently move toward each other. Their
collision yields a peak density (panel at $t=800$ ps). 
Not surprisingly, threads behave as the filaments  described in  Sect. III.B.
A similar pattern of alternating droplets and threads 
was observed in the study of the breakup of inviscid and irrotational capillary jets
discussed in Ref. \onlinecite{Man90}.

The lack of dissipation makes  droplets and threads 
oscillate during the time elapsed by the simulation. Once formed, they
execute a series of vibrations, being alternately compressed and elongated in the
jet direction  with an expected frequency of the order of
$ \omega = \sqrt{8\gamma /m\rho_0 R_0^3}$.\cite{Ray79b}
It has been pointed out that no obvious effects due to superfluidity
have been observed on the breakup behavior of a liquid He jet.\cite{Spe20} Yet,
Kolatzki et al.\cite{Kol22} have found  that He droplets undergo  shape oscillations
that persist for much longer times than in the case of viscous drops, a signature of the
superfluid character of these droplets. 

We would like to mention that if only the cylinder density is perturbed
and no axial velocity field is applied to it, we find that
jet breaking proceeds as in Fig. \ref{fig2},  
the only difference being that 
it takes more time for the instability to fully develop and eventually lead  to jet fragmentation.

The actual time taken for the jet to break into droplets depends upon the amplitude
of the initial density perturbation. It is defined as the time $\tau_b$
it takes for the wave amplitude with the largest frequency to grow up to $R_0$\cite{Dum08,Hoe13}
\begin{equation}
R_0=\delta _0 e^{\omega_{max} \tau _{b}}
\end{equation}
where $\omega _{max}$ is given in Eq. (\ref{omm}).
With our choice for the initial perturbation amplitude $\delta _0$
we have $\tau_b= [ln(R_0/\delta _0)]/\omega_{max}= 695 \,{\rm ps}$.

\begin{figure}
    \centering
    \includegraphics[width=1.0\linewidth,clip]{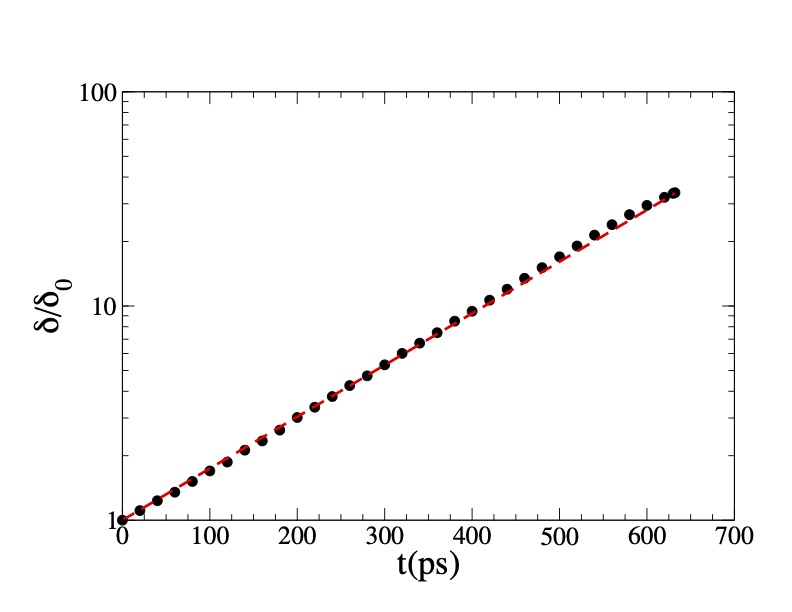}
    \caption{Neck shrinking as a function of time, shown on
a log-scale (see the text for the definitions of $\delta $ and $\delta_0 $).
The points are the numerical values obtained from the 
simulation, whereas the dashed line shows the prediction of linear theory.}
    \label{fig3}
\end{figure}

We have computed the dynamics of neck shrinking by monitoring during the
real-time evolution the quantity
$\delta(t)=(R_{max}-R_{min})/2$, where
the radii $R_{max}$ and $R_{min}$ are measured at the two positions $x=L/4$ and $x=L/2$
(see Fig. \ref{fig2}).
The calculated values for $\delta(t)/\delta_0$ are shown in Fig. \ref{fig3}
on a logarithmic scale  as a function 
of time, and are compared with the quantity 
$e^{\omega _{max}\,t}$ predicted from 
linear theory.
We find remarkable the good agreement, for the whole duration of the 
breaking process, between our simulations and linear theory.

\begin{figure*}
    \centering
    \includegraphics[width=0.9\linewidth,clip]{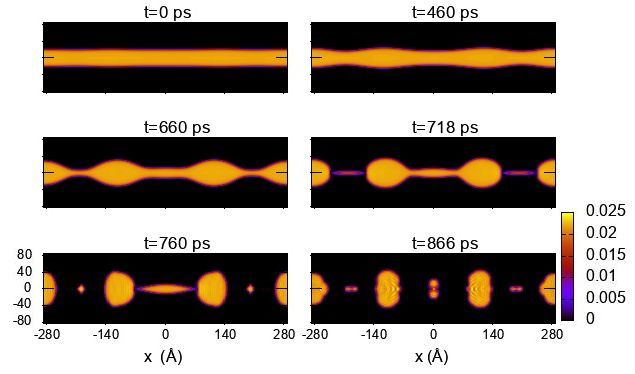}
    \caption{Breaking dynamics of a cylinder subject 
to multiple wavelength axial perturbations as explained in the text.
The color bar shows the atom density in units of \AA$^{-3}$.}
    \label{fig4}
\end{figure*}

Finally, we have also investigated another scenario when 
the jet is subject to a more general perturbation on the equilibrium density, i.e. we have started the real-time dynamics with 
the cylinder simultaneously perturbed by several axisymmetric perturbations of different wavelengths.
In order to accommodate a reasonable number of modes with different wavelengths
compatible with the PBC used here, we  perform the simulation in a cell longer than the one shown in Fig. \ref{fig2}, 
with $L$ equal to three times the critical wavelength associated with the fastest mode, $\lambda_c=2\pi R_0/0.697$. 
We therefore consider an axially symmetric density perturbation given by a linear combination of
six modes with small random amplitudes $\epsilon= \delta_0/R_0$ in the $(-0.03, 0.03)$ range, and wavelengths 
$\lambda_c, 3\lambda_c, 3\lambda_c/2, \lambda_c/2, \lambda_c/4$, and $3\lambda_c/4$,
and perform a real-time simulation starting from such initial state ($t=0$ panel in Fig. \ref{fig4}).

We show in Fig. \ref{fig4} some snapshots of the density 
(on a plane containing the cylinder axis) taken during the real-time evolution 
of this system, where it appears that 
among the various modes, the one eventually dominating in the course of 
time is indeed the critical one, dictated by $\lambda _c$,
which leads to the formation of three necks, eventually resulting
in the fragmentation into three droplets.
However, at variance with the 
case where the critical mode is the only one present (as shown in Fig. \ref{fig2})
the jet does not break up into equal-size droplets.
For much longer filaments than the one investigated here, one might expect
a distribution  of slightly different drop sizes, some drops coming from 
the crests of the primary waves and others from the ligaments linking them.
Determining the disturbance frequencies for jet breaking 
leading to the production of uniformly sized equidistant
He drops has been one of the main concerns of a recent 
work\cite{Kol22} in view of their experimental use
in e.g. coherent diffraction imaging at x-ray free electron lasers. 

We have thus seen that the He-DFT approach is able to address jet breaking yielding results in agreement
with linear theory. This is a needed first step before carrying out the study of contracting $^4$He filaments which we address in the following.

\bigskip
\subsection{Contraction and fragmentation of free-standing filaments} 

Classical fluids jet breaking  may lead not only to  drops but also to filaments.\cite{Egg08,Cas12,Man90}
In the experiments a filament is generated between two drops as a secondary product of breakup.\cite{Wan19}
The appearance of filaments is obviously due to hydrodynamical instabilities developing in the jet but, to the best of our
knowledge, the conditions under which they appear are not fully understood. Their study is further hampered by 
the difficulty  to experimentally generate free cylindrical filaments whose properties could be compared with 
theoretical models describing their dynamical evolution. Only recently it has been possible to carry out a comprehensive  experimental
study of the breakup of normal liquid filaments based on an apparatus originally designed to replicate jet and droplet formation 
in an ink-jet printer, but on a larger scale to produce free cylindrical filaments.\cite{Cas08} 

As with classical fluids, $^4$He jet breaking may lead not only to  droplets but also to 
filaments, as observed in experiments\cite{Spe20,Tan20,Kol22} and we have shown 
that they may appear --not on purpose-- between drops, see Figs. \ref{fig2} and \ref{fig4}.
Here we model these filaments as cylinders of radius $R_0$
delimited by two hemispherical caps\cite{Ant19} and 
study, using the He-TDDFT approach,
their contraction due to the effect of the surface tension
for different values of the
aspect ratio $\Gamma =\tilde{L}/R_0$,\cite{Cas12} where $\tilde {L}$ is the half-length 
of the filament from end-to-end. 
The configuration from which the real-time dynamics is initiated is 
a free-standing ideal filament (i.e. no density perturbation is applied),
as usually done in numerical simulations of the 
contraction of viscous fluid filaments.\cite{Sch96,Ant19,Hoe13}

We have investigated filaments with different values of the aspect ratio,
namely $\Gamma = 4,5,6,8,10.5$, and $15$. Some of these values 
coincide with those studied in Ref. \onlinecite{Ant19} 
at  $Oh=0.001$, which is considered to correspond to the inviscid regime.
Our goal is to study how the initial aspect ratio $\Gamma$ 
determines the fate of the filament, i.e. either contraction into a single 
liquid body (stable state) or breaking into two or more droplets.
Experimentally,\cite{Cas12} it has been found for classical fluids that 
there is a critical initial aspect ratio $\Gamma =6 \pm 1$ below which a liquid filament 
is stable irrespective of the $Oh$ value, and above which the
filaments tend to break into separate droplets.

In the following we describe the most salient features  
found during the
real-time evolution of superfluid $^4$He filaments.
All simulations discussed below
are displayed as movies in the supplementary material accompanying this work.
These movies last for longer times  than those reported in the 
following figures. We do not  discuss the filament appearance  for such 
long times because undamped excitations and especially the annihilation of vortex rings, 
as discussed in the following, tend to produce turbulence\cite{Esc19,Pi21}  
whose description is beyond the scope of this paper.

\subsubsection{Filament with $\Gamma =4$}

This is the shortest filament that we have investigated. Its time evolution 
is similar to that predicted for short filaments by
classical calculations and experiments,\cite{Ant19,Cas12} i.e. the filament
contracts and oscillates back and forth without breaking. 
In the presence of some viscosity, the final configuration would be a 
single spherical droplet.

\begin{figure*}
    \centering
    \includegraphics[width=0.9\linewidth,clip]{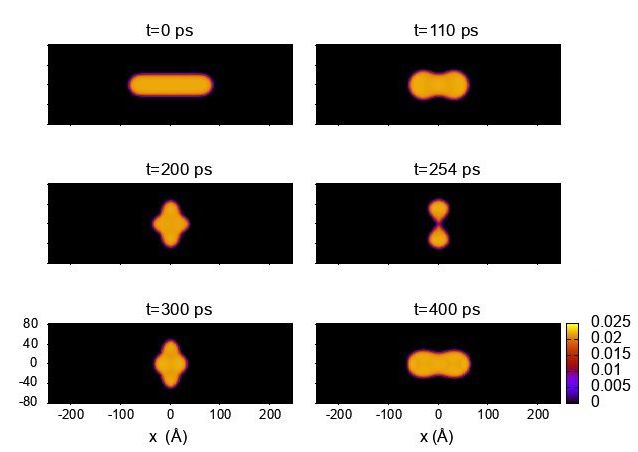}
    \caption{Contraction of a filament with aspect ratio $\Gamma =4$.
    The color bar shows the atom density in units of \AA$^{-3}$.}
    \label{fig5}
\end{figure*}

As shown by the temporal sequences in Fig. \ref{fig5},
a blood-cell shape develops in the transverse direction ($y-z$ plane) 
which is clearly visible at $t=254$ ps, displaying 
an almost empty hole in the center (toroidal shape, we recall that the system is axially symmetric around the $x$ axis)  
before becoming compact again (frame at $t=300$ ps), and
recovering a peanut-like shape along the filament axis
(frame at $t=400$ ps).
For larger times (not shown in the Figure but visible in the movies 
in the supplementary material), after  
extending transversally   
it is drawn again into a compact droplet,
originating a high density spot in its center.
This high density spot relaxes 
and launches a series of density waves 
propagating inside the filament. This effect has been  observed in previous simulations 
of the merging of two $^4$He nanodroplets.\cite{Esc19,Pi21}

\subsubsection{Filament with $\Gamma =5$}

\begin{figure*}
    \centering
    \includegraphics[width=0.9\linewidth,clip]{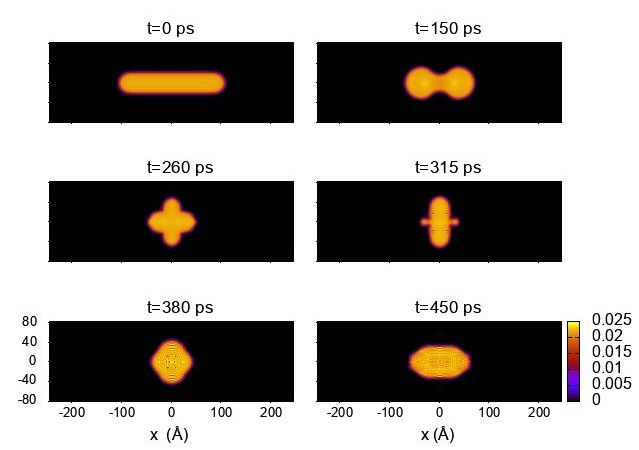}
    \caption{Contraction of a filament with aspect ratio $\Gamma =5$.
    The color bar shows the atom density in units of \AA$^{-3}$.}
    \label{fig6}
\end{figure*}

According to classical calculations and experiments, a filament with this value of $\Gamma$ 
is also expected to display a stable dynamics, 
oscillating back and forth without breaking.\cite{Ant19,Cas12} 
Interestingly, simulation of this filament has disclosed the nucleation of quantized vortex 
rings, which appear in Fig. \ref{fig6} as dark spots in the snapshots at $t=315$ ps and 
$t=380$ ps. For symmetry reasons, only pairs of quantized vortex-antivortex  
rings (vortex ring pairs with opposite circulation) can be nucleated.
No such rings have been found
in classical simulations carried out in the range of Ohnesorge numbers
corresponding to the inviscid regime (below $\sim 2 \times 10^{-3}$). 
Yet, Hoepffner and Par\'e have found classical vortex rings 
for Ohnesorge numbers in the $0.002 < Oh < 0.1$ range
but surprisingly enough 
not in the inviscid regime.\cite{Hoe13} 
We will highlight the role of these vortices 
for the longer filaments 
discussed in the following, where they become effective
in preventing the filament breaking.

In the case displayed in Fig. \ref{fig6},
vortices are nucleated during the
contraction dynamics at surface indentations appearing between 
the end droplets or blobs and the rest of the cylindric filament
(see the panel at $t=315$ ps); this 
requires some inertia which can only be acquired when the aspect ratio of the filament
is larger than a critical value. 
Since this is the calculated filament with smallest $\Gamma$ 
value for which we see vortex rings nucleation,
one should expect the appearance of vortex rings for filaments with  $\Gamma \geq 5$. 
Once nucleated, vortices move to the bulk of the filament. 

The filament end caps  collapse (panel at $t=315$ ps) and launch additional vortex rings. 
One may see multiple vortex-antivortex ring pairs 
in a small volume which eventually 
annihilate, yielding an intense burst of density waves at later times (panel at $t=380$ ps).
Eventually, the filament oscillates between 
the longitudinal and transverse directions, filled with 
density waves propagating inside the formed droplet (panel at $t=450$ ps).
We find similarities with the $L_0 =5$ case in Ref. \onlinecite{Ant19}, 
but at variance with that reference, where 
breakup appears by complex oscillations at $t=5.2\, \tau_c$, 
in our simulation the end caps are reabsorbed in the bulk of the resulting stable droplet.

\subsubsection{Filament with $\Gamma =6$}

\begin{figure*}
    \centering
    \includegraphics[width=0.9\linewidth,clip]{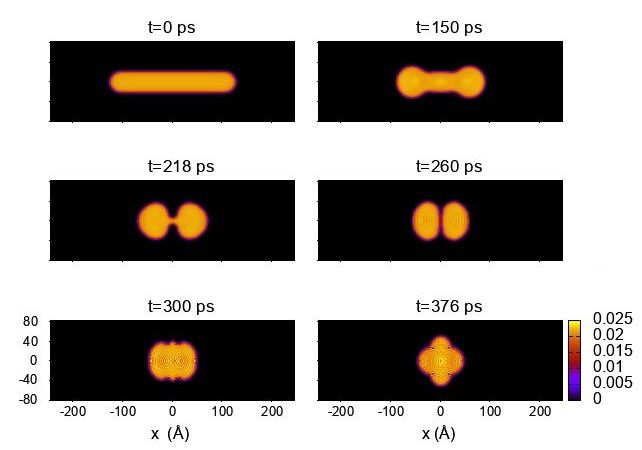}
    \caption{Contraction of a filament with aspect ratio $\Gamma =6$.
    The color bar shows the atom density in units of \AA$^{-3}$.}
    \label{fig7}
\end{figure*}

As shown in Fig. \ref{fig7}, the filament retracts and two end drops appear at the tips, clearly visible 
at around $t= 150$ ps. 
Drops grow in size and the filament between them
contracts and shrinks into a thread, see e.g. the configuration at $t=218$ ps. 
This thread collapses at $t=236$ ps, and the two droplets are temporarily apart,
as shown in the panel for $t=260$ ps.
However, due to the kinetic energy gained during the 
previous contraction stage,    
the two highly deformed fragments collide immediately after and 
merge again at $t \sim 280$ ps to produce a single deformed droplet.
 
The collision of the fragments produces a high density spot at the
contact region (between $t=262$ ps and $t=272$ ps, see movie in the supplementary material) 
which expands yielding density 
waves propagating inside the filament, see the $t=300$ ps frame in Fig. \ref{fig7}.
The merged drop   presents surface 
indentations as those appearing e.g. at $t=300$ ps. These indentations  act as nucleation
sites for quantized vortex rings, which remain close to the 
droplet surface. The cores of some of these vortices are clearly visible in the 
frame at $t=376$ ps. 
Notice that these vortices do not contribute to the escape 
from end-pinching since the thread connecting the end drops has
collapsed before. The density is no longer smooth; rather, it is  
strongly perturbed by the presence of density waves produced by the merging of the two fragments. 

The evolution of this filament is similar to the  
$L_0=6.0$ filament shown in Fig. 5 of Ref. \onlinecite{Ant19}.
For superfluid $^4$He we have found that the filament temporarily 
breaks into two deformed drops  at $t=3.8 \,\tau_c$, similar to the value 
one can read  in Fig. 5 of that reference. 
However, in our case drops collide and merge again, whereas in Ref. \onlinecite{Ant19} they
seem to remain separated.
Another difference between classical and superfluid filaments is the appearance of quantized vortex rings
and their subsequent annihilation.

\subsubsection{Filament with $\Gamma =8$}

\begin{figure*}
    \centering
    \includegraphics[width=0.9\linewidth,clip]{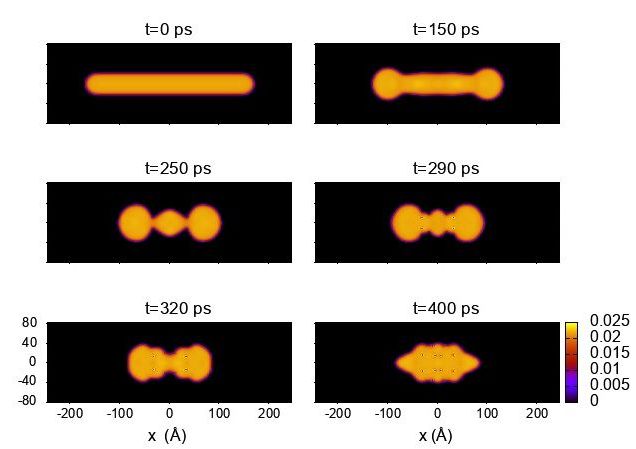}
    \caption{Contraction of a filament with aspect ratio $\Gamma =8$.
    The color bar shows the atom density in units of \AA$^{-3}$.}
    \label{fig8}
\end{figure*}

\begin{figure}
    \centering
    \includegraphics[width=1.0\linewidth,clip]{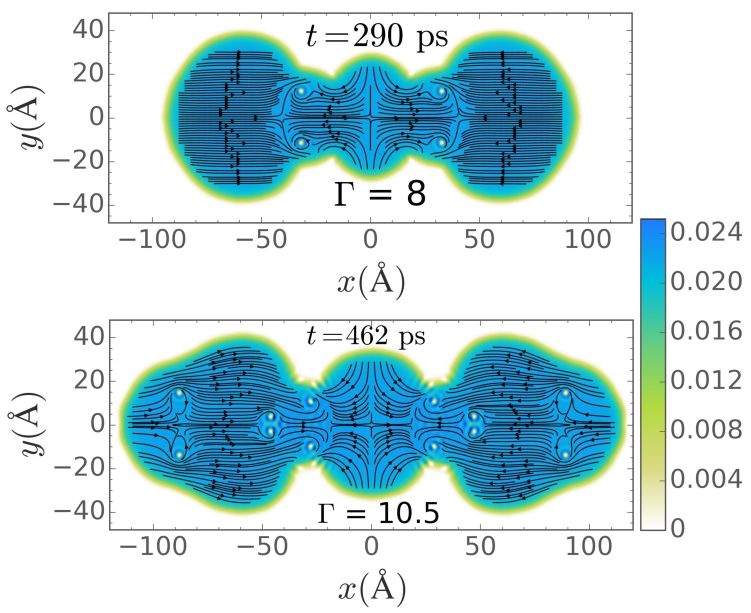}
    \caption{Superfluid streamlines corresponding to the configurations 
    $\Gamma= 8$ at $t=290$ ps (top), and $\Gamma=10.5$ at $t=462$ ps (bottom).
    The color bar shows the atom density in units of \AA$^{-3}$.}
    \label{fig9}
\end{figure}

The dynamical evolution of this filament is shown in Fig. \ref{fig8}.
As for the previous case, end drops develop, 
clearly visible already after $t \sim 100$ ps. 
The main filament connecting the end drops 
shrinks and a thin neck develops at the drop-filament 
contact region, which start pinching off the filament 
with two necks that reach their smallest radius at $t=254$ ps.
Before they completely shrink, vortex rings 
nucleate close to the necks at about $t=258$ ps, being clearly formed at $t=270$ ps.
The streamlines of the superflow are drawn in the top panel of Fig. \ref{fig9} for the configuration at $t=290$ ps,
clearly showing the characteristic pattern of lines wrapping the vortex core positions.

These vortex rings prevent  necks  from 
pinching, as they reopen immediately after their appearance 
(see e.g. the frame at $t=290$ ps), similarly to the mechanism 
discussed by Hoepffner and Par\'e.\cite{Hoe13} 
A flow through the neck develops because of the retraction and, according to these authors,
this flow may detach into the jet downstream of the neck when fluid 
viscosity exceeds a threshold ($Oh \gtrsim 2 \times 10^{-3}$);\cite{Hoe13}  this sudden detachment creates a 
vortex ring which strongly modifies the flow pressure: fluid is 
transported back into the neck which in turn reopens. It is remarkable 
that the same happens in the case of superfluid $^4$He in spite
of the lack of viscosity.
At $t=330$ ps, another pair of vortex rings is nucleated at  
the droplet-filament indentation preventing pinching again. 
Finally,  vortex-antivortex rings annihilate and  disappear from the system
producing as a result a burst of density waves.

The movie in the supplementary information shows the appearance of  surface protrusions
at $t=452$ ps  which act as vortex nucleation sites, and their collapse yields a high density spot.
Eventually, the contracted filament is permeated by a large number of vortex rings at 
$t=488$ ps. This is at variance with the classical, inviscid fluid description.

The evolution of this filament can be compared
to that corresponding to $L_0=8.0$ shown in Fig. 5 of Ref. \onlinecite{Ant19}.
Besides the vortex rings phenomenology, which is absent
in the simulations of that reference, in our
case end-pinching (i.e., the detachment of two droplets from the two ends of a retracting filament)
strictly never happens. The closest the $^4$He filament gets to it is at $t= 4.1 \,\tau_c$, 
whereas the time for the filament  breakup by end-pinching read from Fig. 5 of Ref. \onlinecite{Ant19}
is $t \sim 4.6 \,\tau_c$.

\subsubsection{Filament with $\Gamma =10.5$}

\begin{figure*}
    \centering
    \includegraphics[width=0.9\linewidth,clip]{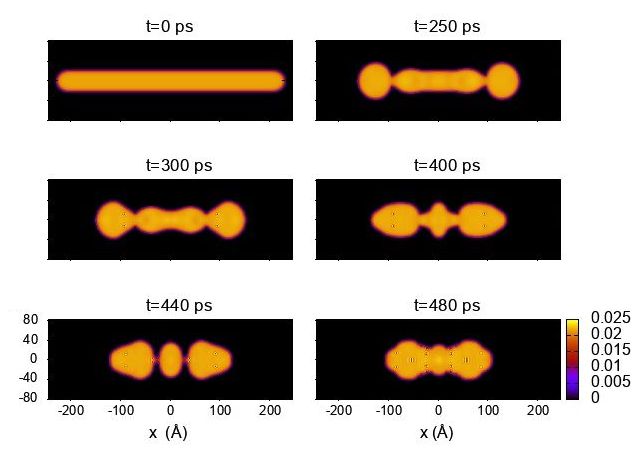}
    \caption{Contraction of a filament with aspect ratio $\Gamma =10.5$.
    The color bar shows the atom density in units of \AA$^{-3}$.}
    \label{fig10}
\end{figure*}

Similarly to the previous cases, end drops develop as shown in Fig. \ref{fig10}.
A more violent approach is 
expected because the filament is longer and end drops have more time to accelerate
under the traction exerted by surface tension.
The filament connecting the end drops contracts 
and necks appear at the drop-filament contact 
region, as shown at $t=250$ ps, which start pinching. 
The neck shrinks to a minimum at $t=256$ ps, escaping from end-pinching again
because vortex rings are nucleated  at $t \sim 260$ ps.

Vortex rings detach from the neck and move towards 
the bulk of the end drops  (frame at $t = 300$ ps).
The remaining filament develops bulges, which evolve to 
a more complex structure (frame at $t=400$ ps).

The snapshot at $t=440$ ps shows an almost complete fragmentation. However, due to the
opposite velocities acquired during the early stages of the contraction, 
the three fragments merge again. Other vortex rings are created in the process, 
nucleated at the necks during the re-merging, as shown in 
the frame at $t=480$ ps. 
The streamlines of the superflow are drawn in the 
bottom panel of Fig. \ref{fig9} for the configuration at $t=462$ ps.

Vortex ring annihilation at later times (see movie in 
the supplementary material) produces density waves arising from the collapse of their cores.
This is a phenomenon that we have not observed in the merging of $^4$He droplets,\cite{Esc19,Pi21} 
nor the shrinking of a vortex ring up to it collapses. It is interesting to see that these small vortex 
rings travel towards the tips of the filament, evaporating from them. 
Eventually, vortex rings disappear and the contracted filament 
enters a complex dynamic regime, hosting plenty of density waves until the end of the simulation. 

The evolution of this filament should be similar to that corresponding 
to $L_0=10.0$ shown in Fig. 5 of Ref. \onlinecite{Ant19}.
Besides the vortex rings phenomenology and wave dynamics, in our
case end-pinching 
strictly never occurs. The filament gets close to it  
at $t= 4.1\,\tau_c$ (254 ps) and especially at $t= 7.0 \,\tau_c$ (432 ps),
whereas the  breakup time  by end pinching read  from Fig. 5  of Ref. \onlinecite{Ant19}
is $t \sim 4.8\,\tau_c$.

\subsubsection{Filament with $\Gamma =15$}

\begin{figure*}
    \centering
    \includegraphics[width=0.9\linewidth,clip]{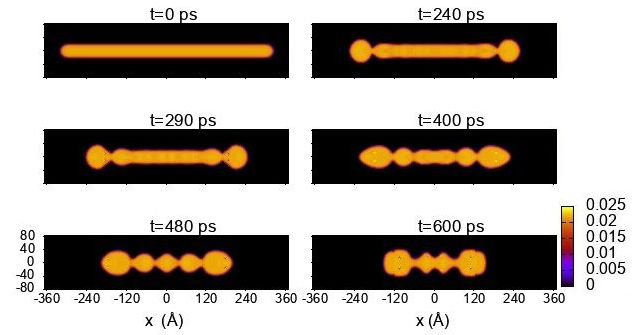}
    \caption{Contraction of a filament with aspect ratio $\Gamma =15$.
    The color bar shows the atom density in units of \AA$^{-3}$.}
    \label{fig11}
\end{figure*}

This is the largest filament we have investigated. 
In classical simulations of sufficiently long filaments 
and small $Oh$ numbers,
as the filament contracts it will succumb to end pinching\cite{Sto86,Sch96,Cas12}
even in cases where the Rayleigh-Plateau instability  
is expected to develop, subsequently resulting in the filament to break up into several drops.
However, this instability does not occur,
suggesting that
the timescale for the  Rayleigh-Plateau instability to grow is much larger than
the timescale for the filament to fully contract even for long filaments.
The same happens in the case of $^4$He, as it can be seen from Fig. \ref{fig11}.

In the case of superfluid $^4$He, however, end-pinching is again hindered by 
vortex ring generation.
The temporal sequence is similar to 
the $\Gamma =8$ and $\Gamma =10.5$ cases, except that the number of necks 
has increased. Well developed end drops appear at $t=100$ ps, with a well developed necks at $t=160$ ps. 
Figure \ref{fig11} shows that end drops nearly pinch-off at $t=248$ ps, but at $t=264$ ps one may see 
vortex rings appearing at the necks, hindering end-pinching.
The vortex rings detach from the neck and move towards the bulk of the 
end drops and bulges appear in the filament  close to the end drops (panel at $t=290$ ps). Bulges evolve to bulbs and,
similarly to the $\Gamma =8$ and $\Gamma =10.5$ cases, intermediate drops 
develop during the time evolution whose number increases with the
length of the filament, as also observed in the 
simulations of classical low viscosity ($0.003 \leq Oh \leq 0.02$) filaments.\cite{Wan19}

The evolution of this filament should be compared to that 
corresponding to $L_0=15.0$ shown in Fig. 5 of Ref. \onlinecite{Ant19}.
Besides the phenomenology of vortex rings proliferation, also in this case
end-pinching never occurs. End drops are close to detach at 
$t= 4.02\,\tau_c$ (247 ps) but escape pinch off because of vortex ring 
nucleation, whereas the filament breakup time read from Fig. 5 of that reference  
is $t \sim 4.8\,\tau_c$.

Finally, we have computed the contraction velocity
for all the investigated filaments.
We have defined the position of the tip of the filament as the location of its sharp surface (that at which the density 
equals $\rho_0/2$) on the  $x$-axis. 

\begin{figure}
    \centering
    \includegraphics[width=1.0\linewidth,clip]{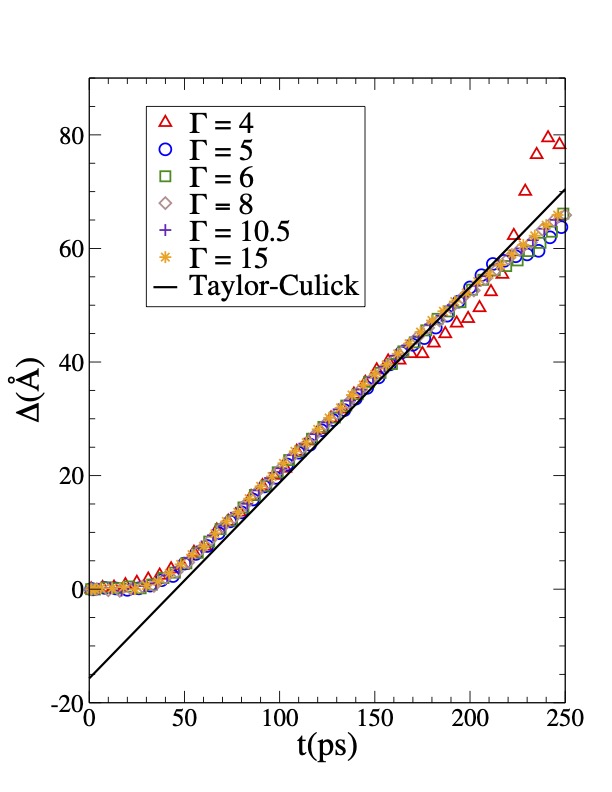}
    \caption{Contraction of filaments with different aspect ratios as a function of time.
The slope of the solid line is the theoretical contraction velocity $R_0/\tau_c$.
}
    \label{fig12}
\end{figure}

Figure \ref{fig12} shows the displacement of the tip position 
as a function of time for the studied filaments. It appears that all curves collapse onto the same curve up to $t \sim 170$ ps (2.76 $\tau_c$).
Consequently,  within this range of time the retracting velocity is independent from the aspect ratio $\Gamma$.
For times in the $50 \,{\rm ps} \leq t \leq 170 \,{\rm ps}$ range,
all filaments accurately follow the line with the slope equal to 
the Taylor-Culick velocity $v=R_0/\tau_c= 0.348$ \AA /ps, 
which is the relevant velocity scale expected for the
retraction process, originally proposed \cite{Tay59,Cul60} as the steady-state velocity
of a capillary-driven retracting inviscid planar liquid where inertia effects balance the capillary forces
acting on the system.
For longer times the behavior changes because there are either filament oscillations, 
changes in the tip shape, or both. The shorter the filament, the earlier these deviations start to show up.
The retracting velocity of liquid filaments has been studied 
for Ohnesorge numbers $Oh \geq 0.1$,\cite{Pie20} finding that the tip dynamics 
is characterized by an oscillating  velocity whose mean value is close to the Taylor-Culick prediction.
These oscillations have also been found for $Oh=0.05$ in the $\Gamma =20$ case.\cite{Pie20} 
In superfluid $^4$He, though, we do not observe any oscillation with time of the tip retraction velocity.

\section{Summary and concluding remarks}

We have studied the instability and breakup of nanoscopic superfluid $^4$He 
jets and filaments within He-DFT at zero temperature. 
We find that the fragmentation
of long cylindrical jets closely follows the predictions of linear theory for inviscid 
fluids, resulting in the formation of larger droplets
intercalated with smaller satellite droplets.

While some of our results for the contraction of free-standing filaments are 
consistent with those
obtained in the inviscid regime which corresponds  
to Ohnesorge numbers smaller than $2 \times 10^{-3}$,\cite{Ant19} 
the novelty with respect to previous calculations 
for classical inviscid filaments is the appearance of
quantized vortex rings in filaments with aspect ratio $\Gamma > 5$. 

Non-quantized vortex ring nucleation  in the region 
connecting the end drops with the rest of the filament plays 
a central role in escaping filament breakup 
in the low-to-intermediate viscosity regime
characterized by Ohnesorge numbers in the  $0.002 < Oh < 0.1$ range.\cite{Hoe13}
Our simulations show that a similar mechanism, associated with quantized vortex
rings, is active in the superfluid regime at zero temperature, mostly preventing the 
droplet formation through end-pinching.
Vortices are also nucleated at surface protrusions appearing in the course of filament 
oscillations, similar to those found in the merging of $^4$He droplets. 
As a result, filaments are permeated by vortex-antivortex ring pairs
whose annihilation yields phonon/roton bursts which may 
leave the filament in a turbulent state.\cite{Esc19,Pi21} 

A key question is why vortex rings, which have appeared in the 
solution of the Navier-Stokes equation in the $0.002 < Oh < 0.1$ regime,  
cease to appear in the inviscid regime\cite{Hoe13,Ant19} whereas we have 
found them in the superfluid regime within the He-DFT
approach. 
It is known that the Gross-Pitaevskii and He-TDDFT equations, appropriated for superfluids, do not 
reduce to the zero-viscosity limit of the Navier-Stokes equation (Euler equation) for a barotropic fluid in irrotational flow.\cite{Bar16} 
In the superfluid case, an extra term appears involving the gradient of the expression 
\begin{equation}
Q=\frac{\hbar^2}{2m} \, \frac{\nabla^2 \rho^{1/2}}{\rho^{1/2}} 
\end{equation}
the so-called quantum pressure term. This term, which is missing in any classical approach, plays
an important role when the density is highly inhomogeneous, as it happens near the core of a quantized vortex.
At variance, it is an ingredient naturally included in the Schr\"odinger He-TDDFT Eq. (\ref{eq3})
and time-dependent Gross-Pitaevskii equation as well.\cite{Bar16}

Notice that the vortices found in this work are vortex rings. These vortices carry no angular momentum and are created in vortex anti-vortex ring pairs. 
We have found in the simulations that  either they annihilate upon vortex-antivortex ring collisions, yielding an intense roton burst, or ``evaporate'' from the filament surface, leaving in both cases vortex-free droplets.
The commonly accepted basic mechanisms for vortex nucleation in $^4$He droplets are friction against the nozzle when exiting droplets still are in the normal 
phase, or  by droplets collisions within the droplet beam at some distance from the nozzle. Capture of impurities by $^4$He droplets may also yield  
vortex rings and vortex loops nucleation, but they are rather short-lived.\cite{Mat14,Lea14}
It is worth noticing the interest in knowing whether droplets formed 
upon breakup of the $^4$He jet carry or not vortices, as their presence affects how nanostructures are assembled within $^4$He droplets.\cite{Ulm23} 

We have shown that the He-DFT approach, which  is a suitable method 
to describe pure and doped superfluid $^4$He nanodroplets, can also address 
superfluid $^4$He jet breaking and the contraction  of superfluid $^4$He filaments. 
The simulations have been carried out  for nanoscopic structures with a $R = 21.5$ \AA{} radius.
The experimental observations, however, have been made on systems 2-3 orders 
of magnitude bigger, so it is natural to wonder if our results
could be related to any experimental observation. 
We want to point out that we have carried out simulations for the $\Gamma=15$ filament using a radius 
twice as larger  and have sensibly found the 
same results.
The crucial point is to simulate structures with a radius large enough so that they have 
developed a bulk region clearly distinguishable from the surface region where the density goes smoothly to zero (finite surface width), as it
happens for the $R$ value we have chosen, see Fig. \ref{fig1}. Once this condition is fulfilled the results
should reflect the behavior of larger, mesoscopic systems 
and the parameter to scale the simulations to larger sizes is, as for classical systems,
the aspect ratio  $\Gamma$.  
The situation is very similar to that of rotating $^4$He droplets, where the experiments have been carried out for droplets 
several orders of magnitude bigger than the simulations\cite{Gom14,Lan18} and, in spite of this, the comparison with simulations
is meaningful.\cite{Anc18} There too, drops were large enough to display distinct bulk and  surface regions.

Yet, our simulations have some unavoidable limitations which prevent from a detailed comparison with experiments. On the one hand,
our approach is strictly a zero temperature approach and there is no dissipation; energy can only be lost by atom evaporation. Due to the 
limited time elapsed by the simulation, at the end of the simulations droplets and filaments still are in an excited state where the system is permeated 
by density waves that should be damped in the long term by any residual viscosity remaining in the system or by atom evaporation.
On the other hand,
we have found that upon filament breaking, the resulting fragments 
have a tendency to merge again into a single droplet. 
Two effects combine to favor this behavior. 
Firstly, fragments, which are nanoscopic, 
have a non-zero surface width that helps recombination due to the overlap of the densities tails.
Secondly, the contraction velocity acquired by the filament in the early stages of the contraction
tends to push together the two highly deformed drops even if they are 
temporarily apart.
One should also consider the role of long-range van der Waals attractive interaction 
between separated fragments, which may also contribute to their merging.
For the much larger sizes in the experiments, however, the vdW forces are expected to be negligible. 
In fact, the force between two spherical particles of diameter $D$ made of $q$ atoms per unit volume 
interacting via the two-body vdW interaction $\lambda/r^6$ is\cite{Ham37} 
$F \propto - \tilde {F}(x)/D$, where $x=d/D$, $d$ being the distance of closest 
approach between the spheres surfaces
and $\tilde {F}(x)\sim -1/(24x^2)$ ($x\ll 1$). 
Therefore, for the sizes encountered in experiments the vdW 
attraction between fragments will be much reduced if
not negligible, meaning that once a filament breaks into two fragments,
recombination into a single droplet due to the vdW attraction is unlikely. 

\section*{SUPPLEMENTARY MATERIAL}   
See supplementary material for the video files showing the real time evolution
of the processes discussed in the present work.

\begin{acknowledgments}
We thank Rico Tanyag for useful exchanges.
This work has been  performed under Grant No.  PID2020-114626GB-I00 from the MICIN/AEI/10.13039/501100011033
and benefitted from COST Action CA21101 ``Confined molecular systems: form a new generation of materials to the stars'' (COSY) 
supported by COST (European Cooperation in Science and Technology).
\end{acknowledgments}

\bigskip

\section*{AUTHOR DECLARATIONS}   
\subsection*{Conflict of Interest}
The authors have no conflicts to disclose.

\subsection*{Author Contributions}
All authors contributed equally to this work.

\subsection*{DATA AVAILABILITY}
The data that support the findings of this study are available
from the corresponding author upon reasonable request

\end{document}